\documentclass[%
 reprint,
showpacs,
 amsmath,amssymb,
 aps,
]{revtex4-1}

\usepackage{graphicx}
\usepackage{dcolumn}
\usepackage{bm}

\newcommand{\tens}[1]{\mbox{\textbf{\textit{\textsf{#1}}}}}
\begin{document}

\preprint{APS/123-QED}

\title{Dynamical Casimir--Polder interaction between a chiral molecule and a surface}

\author{Pablo Barcellona}
\email{pablo.barcellona@physik.uni-freiburg.de}
\affiliation{Physikalisches Institut, Albert-Ludwigs-Universit\"at
Freiburg, Hermann-Herder-Str. 3, 79104 Freiburg, Germany}

\author{Roberto Passante}
\email{roberto.passante@unipa.it}
\affiliation{Dipartimento di Fisica e Chimica,
Universit\`{a} degli Studi di Palermo and CNISM, Via Archirafi 36,
I - 90123 Palermo,
Italy}

\author{Lucia Rizzuto}
\email{lucia.rizzuto@unipa.it}
\affiliation{Dipartimento di Fisica e Chimica,
Universit\`{a} degli Studi di Palermo and CNISM, Via Archirafi 36,
I - 90123 Palermo,
Italy}

\author{Stefan Yoshi Buhmann}
\email{stefan.buhmann@physik.uni-freiburg.de}
\affiliation{Physikalisches Institut, Albert-Ludwigs-Universit\"at
Freiburg, Hermann-Herder-Str. 3, 79104 Freiburg, Germany}
\affiliation{Freiburg Institute for Advanced Studies,
Albert-Ludwigs-Universit\"at Freiburg, Albertstr. 19, 79104 Freiburg,
Germany}

\date{\today}

\begin{abstract}
We develop a dynamical approach to study the Casimir--Polder force between a initially bare molecule and a magnetodielectric body at finite temperature. 
Switching on the
interaction between the molecule and the field at a particular time, 
we study the resulting temporal evolution
of the Casimir--Polder interaction. The dynamical self-dressing of the molecule and its population-induced dynamics are accounted for and discussed.
In particular, we find that the Casimir--Polder force between a chiral molecule and a perfect mirror oscillates in time with a frequency related to the molecular transition frequency, and converges to the static result for large times. 
\end{abstract}

\maketitle

\section{\label{Introduction}Introduction}
Casimir and Casimir--Polder forces (CP) are electromagnetic interactions
between neutral macroscopic bodies and/or molecules due to the  
quantum fluctuations of the electromagnetic field
\cite{Casimir48,CasimirPolder48,Milonni94}. The presence of perfect
boundaries (perfect conductors) modifies the possible wavelength of
vacuum fluctuations and leads to observable effects like 
the lifetime and frequency shift of an atom in an excited state 
\cite{bartonqed}, the interatomic potential between two atoms \cite{power}, the anomalous gyromagnetic
ratio \cite{Kreuzer,bennet}, which are all different from the vacuum case.
 The interaction between a ground-state
molecule and a perfect electric mirror is always attractive while the
interaction between an excited molecule and an perfect electric mirror
shows an oscillating distance-dependence.
This has been confirmed in measurements of the force between an
excited ion and a metallic mirror \cite{wilson,bushev}.

Recently, the attention in the literature has been directed towards chiral molecules which posses distinctive optical properties including optical rotation as well as circular dichroism.
Chiral molecules are molecules without point or plane symmetry; the
two distinct mirror images of a chiral molecule are called
enantiomers. Many of the processes crucial to life involve chiral molecules whose chiral identity plays
a central role in their chemical reactions, the wrong enantiomer reacts in a different way and does not produce the required result.
Spectroscopically, enantiomers have identical properties
and distinguishing between them is a non-trivial task when using normal
spectroscopy. A frequently used method to separate enantiomers in an industrial setting is chiral
chromotography \cite{gil}. Recently, several laser schemes have been proposed to separate mixtures of enantiomers, and the effect of molecular rotation on enantioseparation
has been studied \cite{jacob}. Furthermore Casimir--Polder forces between chiral molecules in absorptive and dispersive chiral medium have shown discriminatory effects, which might be used to separate enantiomers \cite{jenkins,craig,salam,butcher,salam2,salam3}.

In this article, we consider the dynamical Casimir--Polder interaction
between a chiral molecule and a metal or dielectric body
at finite temperature using a dynamical approach, with the molecule exhibiting
electric, magnetic and chiral polarizabilities. 
The dynamical CP force between an enantiomer
and a perfect chiral mirror is a possible system for distinguishing and
separating enantiomers, because the dispersion energy between these
systems depends on the relative handedness of the molecule with
respect to that of the molecules constituting the chiral mirror. 
Therefore, enantiomers that pass at low speeds near the chiral mirror
will be attracted or repelled in opposite directions and will be
separated based on their chirality.

We assume to switch on the
interaction between the molecule and the field at a particular time
and study the resulting time evolution
of the Casimir--Polder interaction.
Even if the interaction with the free field is always present, our assumption to switch on suddenly the interaction with the
body-assisted field at $t_0=0$ can be a good approximation of the more realistic cases of a rapid change of some parameter characterizing the strength of the atom-field interaction or of putting the atom at some distance from the macroscopic body, obtaining a partially dressed atom or molecule \cite{GPPP95}. The
dynamics of the force could be
observed in principle on
time-scales of femto-seconds for typical molecules and nano-seconds for Rydberg atoms. 

The dynamical self-dressing has been considered for an electric ground-state atom near an electric perfect conductor \cite{vasile}, or for a partially dressed
atomic state \cite{messina}. Also, the dynamical CP interaction between a neutral atom and a real surface has been recently investigated \cite{Haakh14}. The CP force
of a chiral molecule near a body has so far been considered only in the static case \cite{butcher}. In this paper, we consider the dynamical self-dressing for a chiral molecule near a body; our approach includes finite temperature, arbitrary geometries of the body and arbitrary internal molecular states.
As a simple application, we consider the CP interaction
between an initially bare ground-state chiral molecule and a perfect
chiral plate at zero temperature. We will show that the
Casimir--Polder interaction at large times can be attractive or
repulsive depending on the chiralities of the molecule and the medium.
This differs from the stationary CP interaction between
an electric molecule and an electric perfect plate, which is always
attractive \cite{CasimirPolder48,barton}. Furthermore, the
time-dependent approach allows us to follow the temporal evolution of
the CP force due to the initial conditions.

The article is organised as follows.
In Sect.~\ref{Sec2}, we consider the Heisenberg dynamics of the
molecule and the body-assisted
field, mutually coupled. Then, in Sect.~\ref{Sec3},
we consider the dynamical Casimir--Polder interaction between a 
molecule with electric, magnetic and chiral responses 
and a body of arbitrary shape
at finite temperature.
In Sect.~\ref{Sec4}, we apply these results to
a particular case: the dynamical CP interaction between
an initially bare chiral ground-state molecule and a perfectly reflecting 
chiral plate at zero temperature.
We close with some conclusion in Sect.~\ref{Sec5}.

\section{\label{Sec2}Dynamics of the molecule and the
body-assisted field}

We consider the mutually coupled temporal
evolution of a single
molecule and the body-assisted
field. The body-field system
is prepared at uniform
temperature $T$, and the molecule in an arbitrary incoherent superposition of internal energy eigenstates. The
dynamics of the molecule can be described with time-dependent flip
operators, defined by
$A_{mn} = \left| m \right\rangle \left\langle n \right|$, where
$\left| n \right\rangle $ is an energy eigenstate.

In order to evaluate the dynamical CP force between the molecule
and the body, we must first solve the molecule-field dynamics to obtain the flip
operators and the field operators in the Heisenberg picture. The total
Hamiltonian is the sum of three terms, the molecule and the field
Hamiltonians and the interaction term in the dipole approximation:
$H = H_A + H_F + H_{AF}$, where 
\begin{align}\nonumber
H_A=& \sum\limits_n E_nA_{nn} \\ \nonumber
H_F =& \sum\limits_{\lambda  = e,m} \int \mathrm{d}^3r
\int\limits_0^\infty  
\mathrm{d}\omega  \hbar \omega \mathbf{f}_\lambda ^\dag \left(
\mathbf{r},\omega  \right) 
\cdot \mathbf{f}_\lambda \left( \mathbf{r},\omega  \right) \\
H_{AF} =&  - \mathbf{d}\cdot \mathbf{E}\left( \mathbf{r}_A \right)
- \mathbf{m}\cdot \mathbf{B}\left( \mathbf{r}_A \right)
\end{align}
where  $\textbf{f}_\lambda \left( \mathbf{r},\omega  \right)$ is the
annihilation operator for the elementary electric and magnetic
excitations of the system \cite{butcher}, $\mathbf{d}$ and $\mathbf{m}$ are respectively the molecule's electric and magnetic dipole moments, and $\textbf{r}_A$
the position of the molecule.

We introduce the Fourier component of the electric field $\mathbf{E}\left( \mathbf{r},\omega
\right)$, $\mathbf{E}\left( \mathbf{r}\right)=\int_0^\infty \text{d}\omega \mathbf{E}\left( \mathbf{r},\omega
\right) +\textup{h.c.} $\\
The commutators between electromagnetic fields read \cite{butcher}:
\begin{align}\nonumber
&\left[ \mathbf{E}\left( \mathbf{r},\omega  \right),\mathbf{E}^\dag
\left( \mathbf{r}',\omega ' \right) \right]  = \frac{\hbar \mu
_0}{\pi } \,\textup{Im}\tens{G}\left( \mathbf{r},\mathbf{r}',\omega
\right)\omega ^2\delta \left( \omega  - \omega ' \right)  \\ \nonumber
&\left[ \mathbf{E}\left( \mathbf{r},\omega  \right),\mathbf{B}^\dag
\left( \mathbf{r}',\omega ' \right) \right]  = \\ \nonumber
& \qquad \qquad -\frac{\mathrm{i}\hbar \mu _0}{\pi }\,\textup{Im}\tens{G}\left(
\mathbf{r},\mathbf{r}',\omega  \right)
 \times \overleftarrow \nabla  '\omega \delta \left(\omega  - \omega '
\right)  \\ \nonumber
&\left[ \mathbf{B}\left( \mathbf{r},\omega  \right),\mathbf{E}^\dag
\left( \mathbf{r}',\omega ' \right) \right] =\\  \nonumber
& \qquad \qquad -\frac{\mathrm{i}\hbar \mu _0}{\pi } \, \nabla  \times
\textup{Im}\tens{G}\left( \mathbf{r},\mathbf{r}',\omega  \right)
\omega \delta \left( \omega  - \omega ' \right) \\ \nonumber
&  \left[ \mathbf{B}\left( \mathbf{r},\omega  \right),\mathbf{B}^\dag
\left( \mathbf{r}',\omega ' \right) \right]  =\\ 
& \qquad \qquad - \frac{\hbar \mu
_0}{\pi } \,\nabla  \times \textup{Im}\tens{G}\left(
\mathbf{r},\mathbf{r}',\omega  \right)
 \times \overleftarrow \nabla  '\delta \left( \omega  - \omega '
\right) 
 \label{comm}
\end{align}
where $\tens{G}$ is the classical Green tensor of the electromagnetic
field and $\left[ \tens{G} \times \overleftarrow \nabla  '
\right]_{ij} = \text{G}_{ik}\varepsilon _{jkl}\overleftarrow \partial
/\partial x'_l$, the Heisenberg equations for the coupled
molecule--field dynamics read:
\begin{align} \nonumber
 &\dot A_{mn} = \mathrm{i}\omega _{mn}A_{mn} + \frac{\mathrm{i}}{\hbar
} \,\mathbf{K}_{mn} \cdot \mathbf{E}\left( \mathbf{r}_A \right) +
\frac{\mathrm{i}}{\hbar } \,\mathbf{Q}_{mn} \cdot \mathbf{B}\left(
\mathbf{r}_A \right) \\ \nonumber
&\dot {\mathbf{E}}\left( \mathbf{r},\omega  \right) =  -\mathrm{i}\omega
\mathbf{E}\left( \mathbf{r},\omega  \right) + \frac{\mathrm{i}\mu
_0}{\pi } \, \omega ^2\textup{Im}\tens{G}\left(
\mathbf{r},\mathbf{r}_A,\omega  \right) \cdot \mathbf{d} \\ \nonumber
& \qquad \qquad+ \frac{\mu _0}{\pi
} \,\omega \textup{Im}\tens{G}\left( \mathbf{r},\mathbf{r}_A,\omega
\right) \times \overleftarrow \nabla  ' \cdot \mathbf{m} \\ \nonumber
&\dot {\mathbf{B}}\left( \mathbf{r},\omega  \right) =  -
\mathrm{i}\omega \mathbf{B}\left( \mathbf{r},\omega  \right) +
\frac{\mu _0}{\pi } \,\omega \nabla  \times \textup{Im}\tens{G}\left(
\mathbf{r},\mathbf{r}_A,\omega  \right) \cdot \mathbf{d} \\  
& \qquad \qquad -\frac{\mathrm{i}\mu _0}{\pi } \,\nabla  \times \textup{Im}\tens{G}\left(
\mathbf{r},\mathbf{r}_A,\omega  \right) \times \overleftarrow \nabla
' \cdot \mathbf{m} 
\end{align}
where $\mathbf{K}_{mn} = \left[ A_{mn},\mathbf{d} \right] $,
$\mathbf{Q}_{mn} = \left[ A_{mn},\mathbf{m} \right] $. 
$\nabla$ and $\overleftarrow \nabla '$ operators act
only on the first and second arguments
of the Green's tensor; for example $\tens{G}\left( \mathbf{r},\mathbf{r}_A,\omega  \right) \times \overleftarrow \nabla  ' = \left. \tens{G}\left( \mathbf{r},\mathbf{r}',\omega  \right) \times \overleftarrow \nabla  ' \right|_{\mathbf{r}' = \mathbf{r}_A}$
Note that an
electric dipole moment can produce a magnetic field and a magnetic
dipole moment can create an electric field; these cross contributions
are the relevant ones for the chiral part of the Casimir force.

In order to include the Lamb shifts and the dissipation of the
molecular system we require the master equations
for the populations $p_n\left( t \right) = \left\langle A_{nn}\left( t
\right) \right\rangle $ and the coherences $\sigma_{nm}\left( t
\right) = \left\langle A_{nm}\left( t \right) \right\rangle $, where
the expectation value is taken over the field thermal state and the
molecular internal state. The evolution of the populations are
governed by the decay rates and the oscillations of the coherences are
governed by the molecular transition
frequencies.  The electric field at the position of the molecule consists
of two terms: the radiation reaction and the free field. As shown in
the literature for a purely electric atom, the radiation reaction field
gives rise to frequency shifts and spontaneous decay for molecule \cite{Buhmann2,ackerhalt}.
We thus renormalise the field by splitting off the radiation
reaction:
\begin{align} \nonumber
 &\left\langle \dot A_{mn} \right\rangle  = \left[ \text{i}\tilde
\omega _{mn} - \left( \Gamma _n +\Gamma _m
\right)/2
\right]\left\langle A_{mn} \right\rangle  \\
& \qquad \qquad+ \frac{\text{i}}{\hbar
} \left\langle \mathbf{K}_{mn} \cdot \mathbf{E}^{\left( 0
\right)}\left( \mathbf{r}_A \right) \right\rangle  +
\frac{\text{i}}{\hbar }\left\langle \mathbf{Q}_{mn} \cdot
\mathbf{B}^{\left( 0 \right)}\left( \mathbf{r}_A \right) \right\rangle
\end{align}
where $m \ne n$ and the expectation value is taken over the field
thermal state and the molecular internal state. $\tilde \omega _{mn}$
are the Lamb-shifted molecular frequencies and $\Gamma_k$ the decay
rates, which have electric, magnetic and chiral contributions. 
Our model hence takes into account the dissipation of the molecular system; in this case there is only
one channel of decay due to the interaction of the molecule with the
bath of electromagnetic modes.

We integrate these equations of motion with respect to the time,
starting from the initial time $t_0=0$ at which the molecule and the
field are uncoupled, to obtain the free and induced flip operator and
electromagnetic fields:
\begin{align} \nonumber
& \left\langle A_{mn}\left( t \right)\right\rangle = \left\langle
A_{mn}^{\left( 0 \right)}\left( t \right)\right\rangle  + \frac{\text{i}}{\hbar }\int\limits_0^t \text{d} t_1f_{mn}\left( t - t_1 \right)\\ \nonumber 
& \quad   \times \left\langle \mathbf{K}_{mn}\left( t_1 \right)
 \cdot \mathbf{E}^{(0)}\left( \mathbf{r}_A,t_1 \right) 
 + \mathbf{Q}_{mn}\left( t_1 \right) \cdot \mathbf{B}^{(0)}\left( \mathbf{r}_A,t_1 \right) \right\rangle \\ \nonumber 
& \mathbf{E}\left( \mathbf{r},\omega  \right) = \mathbf{E}^{\left( 0
\right)}\left( \mathbf{r},\omega  \right) 
+ \sum\limits_{m,n} \int\limits_{0}^t \mathrm{d}t_1 \mathrm{e}^{ -
\mathrm{i}\omega \left( t - t_1 \right)}A_{mn}\left( t_1 \right)  \\
\nonumber 
& \qquad \qquad   \times \left[ \frac{\text{i}\mu _0}{\pi }\,\omega ^2 
\textup{Im}\tens{G}\left( \mathbf{r},\mathbf{r}_A,\omega \right) \cdot \mathbf{d}_{mn} \right.\\ 
\nonumber 
&\qquad \qquad \qquad \quad   \left.  + \frac{\mu _0}{\pi }\,\omega 
\textup{Im}\tens{G}\left( \mathbf{r},\mathbf{r}_A,\omega  \right) 
\times \overleftarrow \nabla  ' \cdot \mathbf{m}_{mn} \right] \\ \nonumber
&\mathbf{B}\left( \mathbf{r},\omega  \right) = \mathbf{B}^{\left( 0
\right)}\left( \mathbf{r},\omega  \right) 
+ \sum\limits_{m,n} \int\limits_{0}^t \mathrm{d}t_1 \mathrm{e}^{ -
\mathrm{i}\omega \left( t - t_1 \right)}A_{mn}\left( t_1 \right) \\ \nonumber
& \qquad \qquad   \times \left[ \frac{\mu _0}{\pi }\,
\omega \nabla  \times \textup{Im}\tens{G}\left( \mathbf{r},\mathbf{r}_A,\omega  \right) \cdot \mathbf{d}_{mn} \right. \\
&\qquad \qquad \qquad \quad \left.  - \frac{\text{i}\mu _0}{\pi }\,
\nabla  \times \textup{Im}\tens{G}\left( \mathbf{r},\mathbf{r}_A,\omega  \right)
 \times \overleftarrow \nabla  ' \cdot \mathbf{m}_{mn} \right]
  \label{eqn1}
\end{align}
where $\textbf{d}_{mn}$, $\textbf{m}_{mn}$
are the matrix elements of the electric and magnetic dipole operators
between the states $\left| m \right\rangle $ and $\left| n \right\rangle $. We will consider time-reversal symmetric systems, where $\textbf{d}_{mn}$ is real and $\textbf{m}_{mn}$ purely imaginary \cite{lloyd} ($\textbf{d}_{mn}=\textbf{d}_{nm}$, $\textbf{m}_{mn}=-\textbf{m}_{nm}$).
Furthermore, we have defined the function:
\begin{equation}
\displaystyle
f_{mn}\left( t \right) = \text{e}^{\left[ \text{i}\tilde \omega_{mn} -
\left( \Gamma _n +\Gamma _m
\right)/2 \right]t}
\end{equation}
The flip operator and the fields are the sum of the free terms, as
they would be in absence of coupling, and induced terms. The molecule
and field systems depend on their history because of their coupling.

\section{\label{Sec3}Dynamical Casimir--Polder force}

We consider the electromagnetic force
due to the interaction of
a molecule exhibiting electric, magnetic and chiral polarisabilities
with the body-assisted field. The field is in
a thermal state with temperature $T$, while the molecule is in a
generic internal state.

\begin{figure}[htbp]
   \centering
   \includegraphics[scale=0.70]{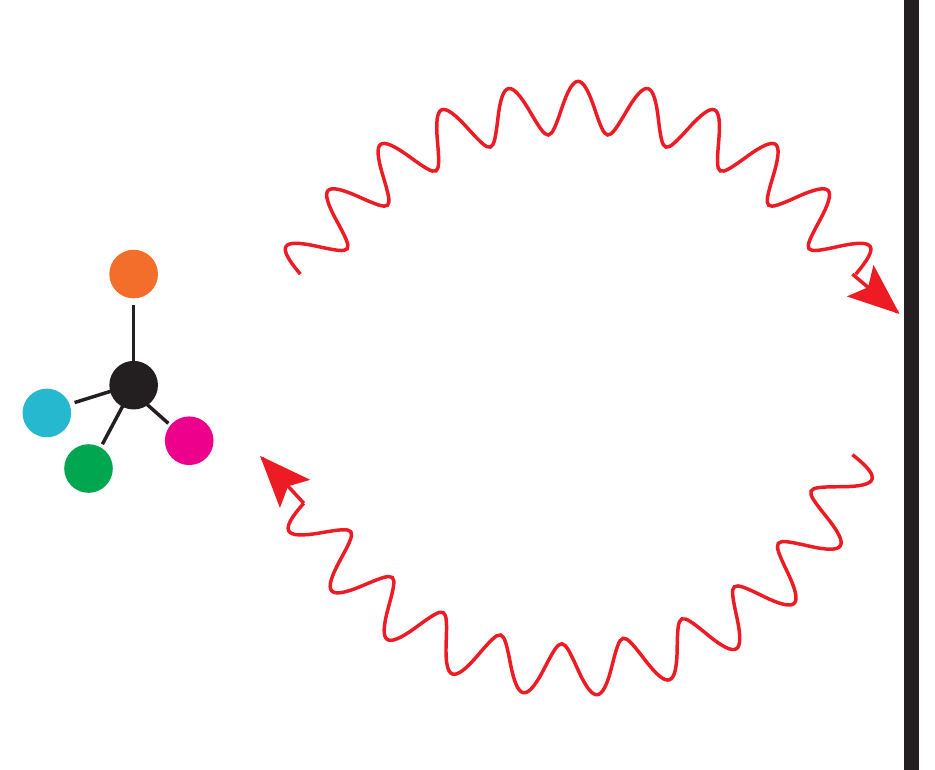}
   \caption{Casimir--Polder force: exchange of a single photon between the chiral molecule and the body.}
    \label{fig}
   \end{figure}

The CP force
between the molecule and the bodi(es)
is due to the exchange of a single
photon: it is emitted, reflected by the bodi(es)
and reabsorbed by the molecule (Fig.~\ref{fig}). The respective
Feynman diagram must contain two interaction vertices which
represent the emission and reabsorption of one photon. The electric contribution
involves two electric-dipole interactions and the magnetic contribution
involves two magnetic-dipole interactions. The chiral interaction involve one
electric-dipole interaction and one magnetic-dipole interaction, in other words:
the interaction must depend on cross terms with one electric dipole
moment and one magnetic dipole moment.

The dynamical CP force for a fixed molecule is:
\begin{multline} 
\mathbf{F} = \left. \nabla \left\langle \mathbf{d}(t) \cdot \mathbf{E}
\left( \mathbf{r},t \right) \right\rangle  \right|_{\mathbf{r} = \mathbf{r}_A} 
+ \left. \nabla \left\langle \mathbf{m} (t) \cdot 
\mathbf{B}\left( \mathbf{r},t \right) \right\rangle  \right|_{\mathbf{r} = \mathbf{r}_A}
\end{multline}
where all operators are obtained by solving the Heisenberg equations and the expectation value is taken over the thermal field state and
the internal molecular state.

We can express the electric field in terms of its free part and
the source field due to the molecule (see Eq.~(\ref{eqn1})):
\begin{multline} 
 \mathbf{F}(t) = \int\limits_0^\infty  \text{d} \omega \sum\limits_{m,n} \left. \nabla \left\langle 
A_{mn}\left( t \right)\mathbf{d}_{mn} \cdot \mathbf{E}^{\left( 0 \right)} 
\left( \mathbf{r},\omega ,t \right) \right\rangle  \right|_{\mathbf{r} = \mathbf{r}_A}\\
  + \int\limits_0^\infty  \text{d} \omega \sum\limits_{m,n} \nabla
   \left. \left\langle A_{mn}\left( t \right)\mathbf{m}_{mn} \cdot
    \mathbf{B}^{\left( 0 \right)} \left( \mathbf{r},\omega ,t \right) \right\rangle
      \right|_{\mathbf{r} = \mathbf{r}_A}   \\
+ \frac{\mathrm{i}\mu _0}{\pi }\sum\limits_{m,n}
\sum\limits_{p,q} \int\limits_0^\infty  \mathrm{d}\omega 
  \int\limits_{0}^t \mathrm{d}t_1 \mathrm{e}^{ - \mathrm{i}\omega
\left( t - t_1 \right)}\left\langle A_{mn}\left( t \right)A_{pq}\left(
t_1 \right) \right\rangle \\ 
 \times \nabla \left\{ \omega
^2 \mathbf{d}_{mn} \cdot \textup{Im}\tens{G}\left(
\mathbf{r}_A,\mathbf{r}_A,\omega  \right)
\cdot \mathbf{d}_{pq} \right. \\
 - \mathbf{m}_{mn} \cdot \nabla  \times
\textup{Im}\tens{G}\left( \mathbf{r}_A,\mathbf{r}_A,\omega 
\right) \times \overleftarrow \nabla  ' \cdot \mathbf{m}_{pq}  \\
 - \mathrm{i}\omega
\mathbf{d}_{mn} \cdot \textup{Im}\tens{G}\left(
\mathbf{r}_A,\mathbf{r}_A,\omega  \right) \times \overleftarrow \nabla
 ' \cdot \mathbf{m}_{pq} \\
  \left.- \mathrm{i}\omega \mathbf{m}_{mn} \cdot \nabla  \times
\textup{Im}\tens{G}\left( \mathbf{r}_A,\mathbf{r}_A,\omega
\right) \cdot \mathbf{d}_{pq}\right\} + \mathrm{c.c.}
\label{eq8}
\end{multline}
As already mentioned, $\nabla$ and $\overleftarrow \nabla '$ operators act
only on the first and second arguments
of the Green's tensor,
respectively:
\begin{align} \nonumber
&\nabla \tens{G}\left( \mathbf{r}_A,\mathbf{r}_A \right) = \left. \nabla \tens{G}\left( \mathbf{r},\mathbf{r}_A\right) \right|_{\mathbf{r} = \mathbf{r}_A} \\
&\nabla '\tens{G}\left( \mathbf{r}_A,\mathbf{r}_A \right) = \left. \nabla' \tens{G}\left( \mathbf{r}_A,\mathbf{r}' \right) \right|_{\mathbf{r}' = \mathbf{r}_A}
\end{align}
In the first two terms in (\ref{eq8}), we use the dynamical equations (\ref{eqn1}) for the flip
operator:
\begin{multline}
 \mathbf{F} (t)= \frac{\text{i}}{\hbar }\int\limits_0^\infty
\text{d}\omega  \int\limits_0^\infty  \text{d}\omega ^\prime
\sum\limits_{m,n} \int\limits_{0}^t \text{d}t_1 f_{mn}\left( t - t_1
\right)  \\ 
 \times \nabla \left\langle \left[ \mathbf{E}^{\left( 0 \right)\dag }
 \left( \mathbf{r}_A,\omega ',t_1 \right) 
 \cdot \mathbf{K}_{mn}\left( t_1 \right) \right. \right.\\
 \left.  + \mathbf{B}^{\left( 0 \right)\dag }
 \left( \mathbf{r}_A,\omega ',t_1 \right)
  \cdot \mathbf{Q}_{mn}\left( t_1 \right) \right]\\
 \left. \left.  \times \left[ \mathbf{d}_{mn}
 \cdot \mathbf{E}^{\left( 0 \right)}\left( \mathbf{r},\omega ,t \right) 
 + \mathbf{m}_{mn} \cdot \mathbf{B}^{\left( 0 \right)}\left( \mathbf{r},\omega ,t \right) \right]
  \right\rangle  \right|_{\mathbf{r} = \mathbf{r}_A}   \\
 + \frac{\mathrm{i}\mu _0}{\pi }\sum\limits_{m,n} \sum\limits_{p,q}
\int\limits_0^\infty  \mathrm{d}\omega 
  \int\limits_{0}^t \mathrm{d}t_1 \mathrm{e}^{ - \mathrm{i}\omega
\left( t - t_1 \right)}
  \left\langle A_{mn}\left( t \right)A_{pq}\left( t_1 \right)
\right\rangle   \\ 
 \times \nabla \left\{ \omega
^2 \mathbf{d}_{mn} \cdot \textup{Im}\tens{G}\left(
\mathbf{r}_A,\mathbf{r}_A,\omega  \right)
\cdot \mathbf{d}_{pq} \right. \\
 - \mathbf{m}_{mn} \cdot \nabla  \times
\textup{Im}\tens{G}\left( \mathbf{r}_A,\mathbf{r}_A,\omega 
\right) \times \overleftarrow \nabla  ' \cdot \mathbf{m}_{pq}  \\
 - \mathrm{i}\omega
\mathbf{d}_{mn} \cdot \textup{Im}\tens{G}\left(
\mathbf{r}_A,\mathbf{r}_A,\omega  \right) \times \overleftarrow \nabla
 ' \cdot \mathbf{m}_{pq} \\
  \left.- \mathrm{i}\omega \mathbf{m}_{mn} \cdot \nabla  \times
\textup{Im}\tens{G}\left( \mathbf{r}_A,\mathbf{r}_A,\omega
\right) \cdot \mathbf{d}_{pq}\right\} + \mathrm{c.c.}
\end{multline}
The thermal expectation value of two positive-frequency electromagnetic
fields is zero.

Next, we use the known formula for the field fluctuations
\cite{butcher}:
\begin{align}\nonumber
&\left\langle \mathbf{E}^{(0)\dag }\left( \mathbf{r},\omega
\right)\mathbf{E}^{(0)}\left( \mathbf{r}',\omega ' \right)
\right\rangle  = \\ \nonumber
& \qquad \qquad  \frac{\hbar \mu _0}{\pi } \,\textup{Im}\tens{G}\left(
\mathbf{r},\mathbf{r}',\omega  \right)\omega ^2\delta \left( \omega  -
\omega ' \right)n\left( \omega  \right) \\ \nonumber
&\left\langle \mathbf{E}^{(0)\dag }\left( \mathbf{r},\omega
\right)\mathbf{B}^{(0)}\left( \mathbf{r}',\omega ' \right)
\right\rangle = \\ \nonumber
& \qquad \qquad    \frac{\mathrm{i}\hbar \mu _0}{\pi
} \,\textup{Im}\tens{G}\left( \mathbf{r},\mathbf{r}',\omega  \right)
 \times \overleftarrow \nabla  '\omega \delta \left(\omega  - \omega '
\right) n\left( \omega  \right)\\ \nonumber
&\left\langle \mathbf{B}^{(0)\dag }\left( \mathbf{r},\omega
\right)\mathbf{E}^{(0)}\left( \mathbf{r}',\omega ' \right)
\right\rangle= \\ \nonumber
& \qquad \qquad   \frac{\mathrm{i}\hbar \mu _0}{\pi } \, \nabla  \times
\textup{Im}\tens{G}\left( \mathbf{r},\mathbf{r}',\omega  \right)
\omega \delta \left( \omega  - \omega ' \right)n\left( \omega  \right)
\\  \nonumber
&\left\langle \mathbf{B}^{(0)\dag }\left( \mathbf{r},\omega
\right)\mathbf{B}^{(0)}\left( \mathbf{r}',\omega ' \right)
\right\rangle = \\
& \qquad \qquad  - \frac{\hbar \mu _0}{\pi } \,\nabla  \times
\textup{Im}\tens{G}\left( \mathbf{r},\mathbf{r}',\omega  \right)
 \times \overleftarrow \nabla  '\delta \left( \omega  - \omega '
\right) n\left( \omega  \right)
\end{align}
where $n\left( \omega  \right)$ is the Bose-Einstein distribution:
\begin{equation}
n\left( \omega  \right) = \frac{1}{\mathrm{e}^{\hbar \omega /k_BT} -
1}\,.
\end{equation}
We also perform the expectation value on the 
inter\-nal molecular state, which is an incoherent
superposit\-ion of energy eigenstates $\left| n \right\rangle $, with probabilities
$p_n$. Furthermore, the two-time correlation function can be
simplified with the Lax regression theorem \cite{onsager,lax} ($t_1
\leq  t$):
\begin{multline}
\left\langle A_{mn}\left( t \right)A_{pq}\left( t_1 \right)
\right\rangle  = f_{mn}\left( t - t_1 \right)\left\langle A_{mn}\left(
t_1 \right)A_{pq}\left( t_1 \right) \right\rangle  =\\ f_{mn}\left( t -
t_1 \right)\delta _{np}\left\langle A_{mq}\left( t_1 \right)
\right\rangle 
\end{multline}
Approximating $\left\langle A_{mn}\left(t_1 \right) \right\rangle  \simeq \text{e}^{\text{i}\tilde \omega_{mn}\left( t_1 - t \right)}\left\langle A_{mn}\left( t \right) \right\rangle $, the CP force is a weighted sum of the CP forces
associated with each eigenstate $
\mathbf{F} (t)= \sum\limits_n p_n(t) \mathbf{F}_n(t) $:
\begin{multline}
\mathbf{F}_n(t) = \frac{\text{i}\mu _0}{\pi }\int\limits_0^\infty
\text{d}\omega  \sum\limits_k \int\limits_{0}^t \text{d}t_1
\text{e}^{ - \text{i}\omega \left( t - t_1 \right)}\\
\times \left\{ f_{nk}\left(
t - t_1 \right)\left[ 1 + n\left( \omega  \right) \right] -
f_{kn}\left( t - t_1 \right)n\left( \omega  \right) \right\} \\
\nonumber
 \times \nabla \left\{ \omega
^2 \mathbf{d}_{nk} \cdot \textup{Im}\tens{G}\left(
\mathbf{r}_A,\mathbf{r}_A,\omega  \right)
\cdot \mathbf{d}_{kn} \right. \\
 - \mathbf{m}_{nk} \cdot \nabla  \times
\textup{Im}\tens{G}\left( \mathbf{r}_A,\mathbf{r}_A,\omega 
\right) \times \overleftarrow \nabla  ' \cdot \mathbf{m}_{kn}  \\
  \left.- 2 \mathrm{i}\omega \mathbf{m}_{nk} \cdot \nabla  \times
\textup{Im}\tens{G}\left( \mathbf{r}_A,\mathbf{r}_A,\omega
\right) \cdot \mathbf{d}_{kn}\right\} + \mathrm{c.c.}
\end{multline}
where $p_n\left( t \right)=\left\langle A_{nn}\left( t \right) \right\rangle $ is the population of 
energy-state $\left| n \right\rangle $. To obtain this expression we have used the Green tensor reciprocity theorem $\tens{G}^T\left( \mathbf{r},\mathbf{r}' \right) = \tens{G}\left( \mathbf{r}',\mathbf{r} \right)$, and for the chiral part the property $\left[ \tens{G}\left( \mathbf{r}_A,\mathbf{r}_A,\omega  \right) \times \overleftarrow \nabla  ' \right]^T$  $=  - \nabla  \times \tens{G}\left( \mathbf{r}_A,\mathbf{r}_A,\omega  \right)$. The term proportional to the Bose-Einstein
distribution describes the interaction between the molecule and the
thermal field, while the term independent of $n$ describes the
interaction with the vacuum field.

Now we introduce $\nabla_A$ which acts on both arguments of the Green tensor.
Exploiting Onsager reciprocity, the relation
\begin{multline}
 \nabla _A \tens{G}\left( \mathbf{r}_A,\mathbf{r}_A \right) \\
= \left. \nabla \tens{G}\left( \mathbf{r},\mathbf{r}_A \right) \right|_{\mathbf{r} = \mathbf{r}_A} 
 + \left. \nabla ' \tens{G}\left( \mathbf{r}_A,\mathbf{r}' \right) \right|_{\mathbf{r}' = \mathbf{r}_A}  \\
= \nabla \tens{G}\left( \mathbf{r}_A,\mathbf{r}_A \right) 
+ \nabla '\tens{G}\left( \mathbf{r}_A,\mathbf{r}_A \right)\\
= \nabla \tens{G}\left( \mathbf{r}_A,\mathbf{r}_A \right) 
+ \nabla \tens{G}^T\left( \mathbf{r}_A,\mathbf{r}_A \right)
\end{multline}
holds. For a time-reversal symmetric molecule, we can hence make the replacement: 
\begin{equation}
\nabla \tens{G}\left( \mathbf{r}_A,\mathbf{r}_A,\omega  \right) \to
\frac{1}{2}\nabla_A \tens{G}\left( \mathbf{r}_A,\mathbf{r}_A,\omega
\right)
\end{equation}
After performing
the time-integrals we obtain:
\begin{align}\nonumber
\mathbf{F}(t) &= \sum\limits_n p_n (t)\left[ \mathbf{F}_n^e(t) + \mathbf{F}_n^m(t) + \mathbf{F}_n^c(t) \right], \\ \nonumber
\mathbf{F}^e_n (t)&= \frac{\mu _0}{2\pi }\int\limits_0^\infty
\text{d}\omega \omega^2  \sum\limits_k \Psi_{kn}(\omega, t)\\ \nonumber
 & \qquad  \times \nabla_A \mathbf{d}_{nk}\textup{Im}\tens{G}\left(
\mathbf{r}_A,\mathbf{r}_A,\omega  \right) \mathbf{d}_{kn},  \\ \nonumber
\mathbf{F}^m_n (t)&=- \frac{\mu _0}{2\pi }\int\limits_0^\infty
\text{d}\omega  \sum\limits_k \Psi_{kn}(\omega, t) \\ \nonumber
 & \qquad  \times  \nabla_A \mathbf{m}_{nk}\nabla  \times \textup{Im}\tens{G}\left(
\mathbf{r}_A,\mathbf{r}_A,\omega  \right) \times \overleftarrow \nabla
'\mathbf{m}_{kn},   \\ \nonumber
\mathbf{F}^c_n (t)&= -\frac{\text{i}\mu _0}{\pi }\int\limits_0^\infty
\text{d}\omega \omega  \sum\limits_k \Psi_{kn}(\omega, t) \\ 
& \qquad  \times   \nabla_A  \mathbf{m}_{nk}\nabla
\times \textup{Im}\tens{G}\left( \mathbf{r}_A,\mathbf{r}_A,\omega
\right)\mathbf{d}_{kn}
\end{align}
where 
\begin{multline}
\Psi_{kn}(\omega, t)= \frac{1 - \text{e}^{ -
\text{i}t\left( \omega  + \omega _{kn}^{\left(  -  \right)}
\right)}}{\omega  + \omega _{kn}^{\left(  -  \right)}}\left[ 1 +
n\left( \omega  \right) \right] \\
- \frac{1 - \text{e}^{ -
\text{i}t\left( \omega  - \omega _{kn}^{\left(  +  \right)}
\right)}}{\omega  - \omega _{kn}^{\left(  +  \right)}}n\left( \omega
\right) +\mathrm{c.c.} 
\end{multline}
and 
$\omega _{kn}^{\left(  \pm  \right)} = \tilde \omega _{kn} \pm
\text{i}\left( \Gamma _n + \Gamma _k \right)/2$. As explained before, for time-reversal symmetric systems, the electric dipole elements are real and the magnetic dipole elements purely imaginary. We have separated the electric, magnetic and chiral contributions: electric contribution contains two electric dipole moments, the magnetic contribution two magnetic dipole moments and chiral contribution contains cross terms with one electric dipole moment and one magnetic dipole moment.

We observe that the force
depends on time in two ways: firstly,
the populations of the internal molecular
states may depend on time. For example, a
molecule initially prepared in some excited state
will unavoidably decay
to the ground state, so the population of the
excited state is one for short times but zero for large times.
The time scale of this population-induced dynamics of the force is
set by the life times \mbox{$\tau_n=1/\Gamma_n$} of the initially
populated states.
For a ground-state molecule the populations of the energy levels are
constant in time and there is no population-induced dynamics.

The time-dependent exponentials on the other hand describe the
dynamical self-dressing of the molecule which is the focus of this work. The self-dressing dynamics operates
on the much shorter time scales of the order of the inverse molecular
transition frequencies $1/\omega_{nk}$. The dynamical self-dressing has been considered for a single non-absorbing electric molecule
in front a plate \cite{vasile}; our approach is generalized for finite temperature, arbitrary geometry of the body and it accounts for molecular absorption.

For times much larger than $1/\omega_{nk}$,
the exponential function is rapidly oscillating and averages to zero. 
The electric part of the CP force then converges to the 
value obtained in the literature with a dynamical approach where the 
time-dependence is solely due to population-induced dynamics
\cite{buh,Buhmann2}.
Our new results hence generalise the previous dynamical approach
to include the self-dressing of the molecule as well as the
magnetic and the chiral parts of the interactions.
The static limit of our result for the
chiral contribution extends previous results from time-independent perturbation theory to
finite temperature and absorbing molecules \cite{butcher}.

As a simple example, we consider an isotropic non-absorbing molecule. The electric, magnetic and chiral parts of the dynamical Casimir--Polder interaction are:
\begin{align} \nonumber
\mathbf{F}_n (t) &=\mathbf{F}_n^e (t) +\mathbf{F}_n^m (t) +\mathbf{F}_n^c(t)  \\ \nonumber
 \quad \mathbf{F}_n^e (t) &= \frac{\mu _0}{3\pi }\int\limits_0^\infty  \mathrm{d}\omega  \omega ^2\sum\limits_k \mathbf{d}_{nk} \cdot \mathbf{d}_{kn} \Psi' _{kn}\left( \omega ,t \right) \\ \nonumber
 &\quad \times  \nabla_A \textup{Tr}\left\{  \textup{Im}\tens{G}\left(
\mathbf{r}_A,\mathbf{r}_A,\omega  \right) \right\},  \\ \nonumber
 \quad  \mathbf{F}_n^m(t) &=  - \frac{\mu _0}{3\pi }\int\limits_0^\infty  \text{d} \omega \sum\limits_k \mathbf{m}_{nk}  \cdot \mathbf{m}_{kn} \Psi' _{kn}\left( \omega ,t \right)\\ \nonumber
 &\quad \times \nabla_A \textup{Tr}\left\{\nabla  \times \textup{Im}\tens{G}\left( \mathbf{r}_A,\mathbf{r}_A,\omega  \right) \times \overleftarrow \nabla  '  \right\}, \\ \nonumber
 \quad \mathbf{F}_n^c (t) &=- \frac{2\mu _0}{3\pi }\int\limits_0^\infty  \mathrm{d}\omega  \omega \sum\limits_k R_{nk}\Psi' _{kn}\left( \omega ,t \right) \\ 
 &\quad \times \nabla_A \textup{Tr} \left\{ \nabla  \times \textup{Im}\tens{G}\left( \mathbf{r}_A,\mathbf{r}_A,\omega  \right) \right\}  
\label{form1}
\end{align}
where 
\begin{multline}
\Psi ' _{kn}\left( \omega ,t \right)  = \frac{1 + n\left( \omega  \right)}{\omega _{kn} + \omega }\left( 1 - \cos \left[ \left( \omega _{kn} + \omega  \right)t \right] \right)\\ +  \frac{n\left( \omega  \right)}{\omega _{kn} - \omega} \left( 1 - \cos \left[ \left( \omega _{kn} - \omega  \right)t \right] \right) ,
\end{multline}
$\textup{Tr}$ is the trace, ${R_{nk}} = \operatorname{Im} \left( {{{\mathbf{d}}_{nk}} \cdot {{\mathbf{m}}_{kn}}} \right)$ is the rotatory strength and $\omega_{kn}$ the transition frequency between the state $\left| k \right\rangle $ and $\left| n \right\rangle $. \\
If either the medium ($\textup{Tr} \left\{ \nabla  \times \textup{Im}\tens{G}\left( \mathbf{r}_A,\mathbf{r}_A,\omega  \right) \right\} = 0$), or the particle ($R_{nk} = 0$)
is achiral there will be no chiral component to the Casimir--Polder potential. This can be
thought of as an application of to the Curie dissymmetry principle (originally formulated for
crystal symmetries): the CP potential cannot distinguish between
molecules of different handedness if the medium does not possess chiral properties itself.

Under reflection, the electric dipole moment changes sign, while the magnetic dipole moment does not. The electric and magnetic parts of the dynamical interaction hence do not change if the molecule is substituted with its enantiomer (mirror image), but the chiral part of the CP force changes sign. This shows the discriminatory effect for the chiral part of the dynamical interaction.

\section{\label{Sec4}Chiral molecule in front a perfect mirror}

The interaction between a ground-state electric mole\-cule and a
perfectly conducting electric plate at zero-temperature has been investigated in the
literature \cite{vasile}; the results can be recovered with our model but we will not focus on
this point here. We consider instead the interaction between a ground-state chiral
molecule and and a perfectly reflecting
chiral plate at zero temperature, $n(\omega)=0$.
As the population of the ground state is constant
in time, the only dynamics of the Casimir--Polder force arises
due to self-dressing.

The Green's tensor of the perfectly reflecting chiral plate
is known, and it depends only in the distance $d$ between the molecule
and the mirror \cite{butcher,Buhmann}:
\begin{multline} 
   \frac{\partial }{\partial d}\left\{ \omega \nabla  \times \textup{Im}\tens{G}\left( \mathbf{r}_A,\mathbf{r}_A,\omega  \right) \right\}= \\
    \pm \frac{3c}{8\pi d^4}\left[ \cos x + x\sin x -
\frac{1}{3} \,x^2\cos x \right]_{x = 2d\omega /c}
\label{eq13}
\end{multline}
where the sign $+$ or $-$ refers to plates of positive and negative chirality, respectively. Note that the trace of
the Green tensor scales differently for small and larges distances
leading to different dependences of the force on the
distance.

After inserting this expression into
Eq.~(\ref{form1}), we next need to perform the frequency integrals for the three terms
in the above Eq.~(\ref{eq13}). This task can be simplified considerably by expressing the Green's
tensor in terms of a differential operator:
\begin{multline} 
   \frac{\partial }{\partial d}\left\{ \omega \nabla  \times \textup{Im}\tens{G}\left( \mathbf{r}_A,\mathbf{r}_A,\omega  \right) \right\} = \\
   \pm \frac{3c}{8\pi d^4} \,\mathop {\lim }\limits_{m \to 1} \left[ 1 -
\frac{\partial }{\partial m} + \frac{1}{3} \frac{\partial ^2}{\partial
m^2} \right]\left. \cos \left( mx \right) \right|_{x = 2d\omega /c}
\end{multline}
Inserting the Green's tensor
in this form,
the chiral interaction reads:
\begin{multline}
 \mathbf{F}^c =  \mp \frac{1}{4\pi ^2\varepsilon _0cd^4} \,
\mathop {\lim }\limits_{m \to 1} \left[ 1 - \frac{\partial }{\partial
m} + \frac{1}{3}\frac{\partial ^2}{\partial m^2} \right]  \\
\times \sum\limits_k  R_{0k} \int\limits_0^\infty \text{d} x \frac{\cos \left( mx \right)}{x + x_k}
\left( 1 - \cos \left[ \left( x + x_k \right)a \right] \right) \hat d 
\end{multline}
where  $x = 2d\omega /c$, $x_k = 2d\omega_k /c$ and $a=ct/(2d)$ and
$\hat d= \textbf{d}/d$.

\subsection{Stationary case: large times}
For times much larger than $1/\omega_k$, 
the cosine function oscillates rapidly and its
contribution vanishes; this situation corresponds to a totally dressed
molecule.

We introduce the auxiliary functions, for $m,y>0$:
\begin{align} \nonumber
&\text{F}\left( m,y \right) = \int\limits_0^\infty \mathrm{d}x
\frac{\sin \left( mx \right)}{x + y}  \\ \nonumber 
&\qquad  =\sin \left( my \right)\operatorname{Ci}\left(
my \right) - \cos \left( my \right)\left[ \operatorname{Si} \left( my \right) -
\frac{\pi }{2} \right] \\ \nonumber
&\text{G}\left( m,y \right) = \int\limits_0^\infty \mathrm{d}x
\frac{\cos \left( mx \right)}{x + y}  \\
&\qquad =  - \cos \left( my
\right)\operatorname{Ci}\left( my \right) - \sin \left( my \right)\left[ \operatorname{Si} \left( my
\right) - \frac{\pi }{2} \right]
\end{align}
where $\operatorname{Si}$ and $\operatorname{Ci}$ are the sine and cosine integral functions.
For large times the CP force converges to the following
static force:
\begin{multline}\nonumber 
\mathbf{F}^c_{t \to \infty } =\mp \frac{1}{4\pi ^2\varepsilon
_0cd^4} \, \sum\limits_k R_{0k}\\
\times \mathop {\lim }\limits_{m \to 1} \left[ 1 - \frac{\partial
}{\partial m} + \frac{1}{3}\frac{\partial ^2}{\partial m^2}
\right]\text{G}\left( m,x_k \right) \hat d \\ 
= \mp \frac{1}{3\pi ^2\varepsilon _0cd^4}\sum\limits_k R_{0k}\\
\times \left[ 1 - 2\operatorname{Ci}\left( 2k_kd \right)f\left( k_kd \right) + \left(
2\operatorname{Si}\left( 2k_kd \right) - \pi  \right)g\left( k_kd \right) \right]
\hat d 
\end{multline}
where $k_k = \frac{\omega _k}{c}$ is the molecular wave number and we
have introduced the auxiliary functions:
\begin{align} \nonumber 
&f\left( x \right) = \frac{3x}{4} \,\sin \left( 2x \right) + \left(
\frac{3}{8} - \frac{x^2}{2} \right)\cos \left( 2x \right) \\
& g\left( x \right) = \frac{3x}{4} \, \cos \left( 2x \right) - \left(
\frac{3}{8} - \frac{x^2}{2} \right)\sin \left( 2x \right)
\label{fg}
\end{align}
This is an alternative, slightly more explicit form for the result
known in the literature \cite{Buhmann,butcher}.

\begin{figure}[htbp]
   \centering
   \includegraphics[scale=0.65]{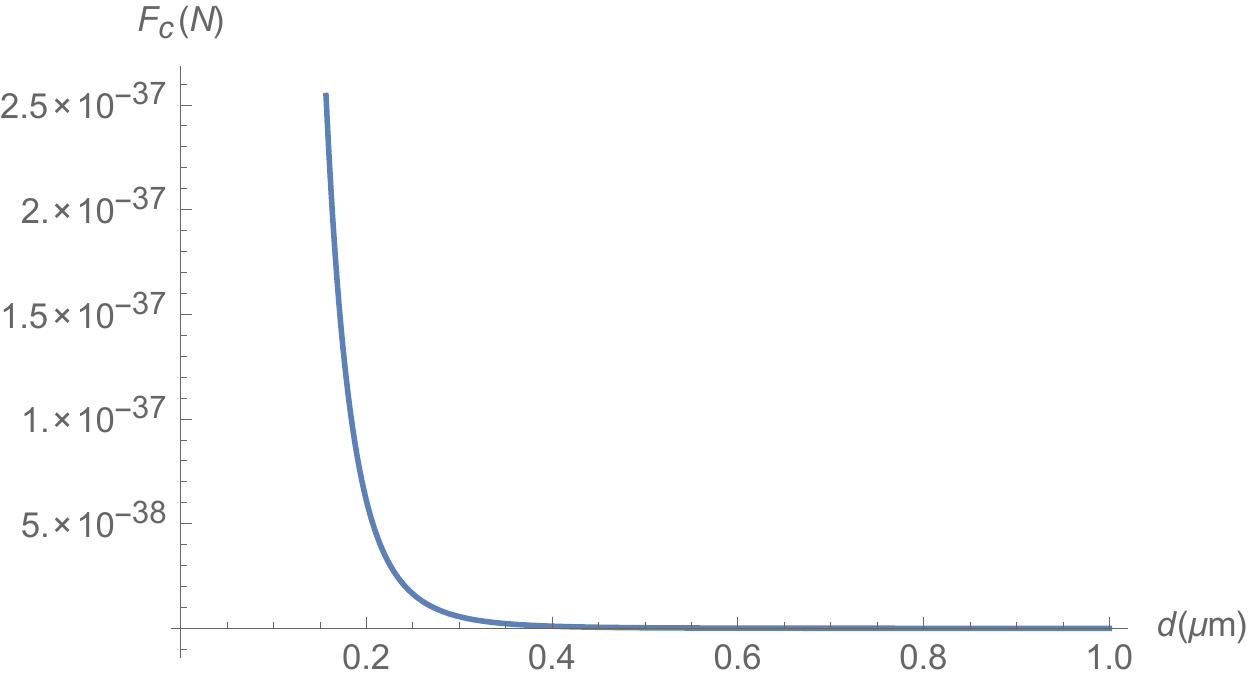}
   \caption{Stationary chiral Casimir--Polder interaction between a
ground-state dimethyl disulphide and a perfect mirror
of negative chirality.}
   \label{fig1}
   \end{figure}

As an example of a chiral molecule, consider
dimethyl disulphide $(\mathrm{CH}_3)_2\mathrm{S}_2$.
The dipole and rotatory strengths for each transition have been
numerically calculated for various orientations \cite{rauk}. As an
example, we have chosen the first transition when the
orientation between the two $\mathrm{CH}_3-\mathrm{S}-S$ planes is
$90^\circ$. The transition frequency between the excited state and the
ground state is
$\omega_{10} = 9.17 \cdot 10^{15} \mathrm{Hz}$, the square of the
dipole moment $\left| \mathbf{d}_{01} \right|^2 = 8.264 \cdot 10^{ -
60}\left( \mathrm{Cm} \right)^2$ and the rotatory
strength is $R_{10} = 3.328 \cdot 10^{-64}
\mathrm{C}^2\mathrm{m}^3\mathrm{s}^{-1}$.
Fig.~\ref{fig1} shows the chiral Casimir--Polder force for a
ground-state dimethyl disulphide molecule above a perfect
mirror of negative chirality.
The stationary chiral CP force between the ground-state
molecule and the medium is repulsive
due to the chosen opposite chiralities of molecule and mirror.
This differs from the CP interaction between an
electric molecule and a perfectly conducting electric plate, which is
attractive \cite{CasimirPolder48,barton,vasile}.

The distance-dependence of the chiral CP force
can be reduced to simple power laws in the
retarded and non-retarded limits. In the non-retarded limit $d
\ll \lambda_k$ or equivalently ${x_k} \ll 1$, we may approximate
$\mathrm{G}\left( m,x_k \right) \to  - \gamma  - \log \left( mx_k
\right)$ where $\gamma$ is the Euler--Mascheroni constant. The force
is then \cite{butcher,Buhmann}:
\begin{equation}\mathbf{F}_{non - ret} = \pm \frac{1}{4\pi
^2\varepsilon _0cd^4}\sum\limits_k R_{0k}\log \left( \frac{\omega
_kd}{c} \right) \hat d \end{equation}

In the retarded limit $d \gg \lambda_k$ or equivalently $x_k \gg 1$, we have
$\mathrm{G}\left( m,x_k \right) \to \frac{1}{\left(
mx_k
\right)^2}$ and the force is \cite{butcher,Buhmann}:
\begin{equation}\mathbf{F}_{ret} = \mp \frac{5c}{16\pi ^2\varepsilon
_0d^6}\sum\limits_k \frac{R_{0k}}{\omega _k^2} \, \hat d
\end{equation}
As expected, the retarded interaction decreases more rapidly due to
the finite velocity of the light: during the time in which the virtual photon has been exchanged, the
molecule will evolve. This associated loss of correlation leads to a more rapidly decreasing force.

Due to the unusual $\log \left( \frac{\omega _kd}{c} \right)/d^4$
dependence of the force in the non-retarded limit, the chiral
potential grows more rapidly than the electric and magnetic
potentials when approaching the surface.

The chiral force changes sign if the molecule is substitute with its
entaniomer, or when the plate of negative chirality
is substituted with one of
negative chirality. 
This discriminatory effect is also observed in the dynamic case, which we
will consider in the next section.

\subsection{Dynamical case}
We now consider the dynamical situation in which the interaction with the perfect chiral
plate starts at the initial time $t_0=0$ and we ask for the
dynamical  Casimir--Polder force between the chiral molecule and the
mirror. In this case, the bare molecular state is not an eigenstate of
the total Hamiltonian, and thus it evolves in time (dynamical
self-dressing), yielding a time-dependent force between the mirror and
the molecule.

To evaluate the force, we use the trigonometric relation
\begin{multline} \nonumber
\cos \left( mx \right)\left( 1 - \cos \left[ \left( x + x_k \right)a
\right] \right)  = \cos \left( mx \right)\\
 - \frac{\cos \left( ax_k
\right)}{2} \,\left\{ \cos \left[ \left( m + a \right)x \right] + \cos
\left[ \left( m - a \right)x \right] \right\} \\
  + \frac{\sin \left( ax_k \right)}{2} \,\left\{ \sin \left[ \left( m + a
\right)x \right] - \sin \left[ \left( m - a \right)x \right] \right\}
\end{multline}
The force has two different expressions before and after the
back-reaction time ($t=2d/c$), which is the time needed for 
light emitted by the atom to
be reflected by the mirror
and return to the molecule.
For  $t<2d/c$ and $t>2d/c$ the chiral dynamical CP force is:
\begin{widetext}
\begin{align} \nonumber
\mathbf{F}^c_{t < 2d/c} &=\mp \frac{1}{4\pi ^2\varepsilon
_0cd^4}\sum\limits_k R_{0k} \mathop {\lim }\limits_{m \to 1} \left[ 1
- \frac{\partial }{\partial m} + \frac{1}{3}\frac{\partial
^2}{\partial m^2} \right]  \left\{ \mathrm{G}\left( m,x_k \right)
+  \right.  \\ \nonumber
& \qquad \left.  - \frac{\cos \left( ax_k \right)}{2} \, \left[ \mathrm{G}\left( m
+ a,x_k \right) + \mathrm{G}\left( m - a,x_k \right) \right] +
\frac{\sin \left( ax_k \right)}{2} \, \left[ \mathrm{F}\left( m + a,x_k
\right) - \mathrm{F}\left( m - a,x_k \right) \right] \right\} \hat d
\\ \nonumber
&=  \mp \frac{1}{3\pi ^2\varepsilon _0cd^4}\sum\limits_k R_{0k} \left[
1 - \frac{8d^4\cos \left( \omega _kt \right)}{\left( c^2t^2 - 4d^2
\right)^2} \right. + \frac{2d^2\cos \left( \omega _kt \right) +
d^2\omega _kt\sin \left( \omega _kt \right)}{c^2t^2 - 4d^2} \\
\nonumber
& \qquad - \left[ 2\operatorname{Ci}\left( 2k_kd \right) - \operatorname{Ci}\left( 2k_kd - \omega
_kt \right) - \operatorname{Ci}\left( 2k_kd + \omega _kt \right) \right]f\left( k_kd
\right) \\ 
& \qquad \left.  + \left[ 2\operatorname{Si}\left( 2k_kd \right) - \operatorname{Si}\left( 2k_kd -
\omega _kt \right) - \operatorname{Si}\left( 2k_kd + \omega _kt \right)
\right]g\left( k_kd \right) \right]\hat d
\end{align}
\begin{align} \nonumber
\mathbf{F}^c_{t > 2d/c} &=\mp \frac{1}{4\pi ^2\varepsilon
_0cd^4}\sum\limits_k R_{0k} \mathop {\lim }\limits_{m \to 1} \left[ 1
- \frac{\partial }{\partial m} + \frac{1}{3}\frac{\partial
^2}{\partial m^2} \right] \left\{  \mathrm{G}\left( m,x_k \right)
+  \right.  \\ \nonumber
&\qquad \left.  - \frac{\cos \left( ax_k \right)}{2} \, \left[ \mathrm{G}\left(
a+m,x_k \right) + \mathrm{G}\left( a-m,x_k \right) \right] +
\frac{\sin \left( ax_k \right)}{2} \, \left[ \mathrm{F}\left( a+m,x_k
\right) + \mathrm{F}\left( a-m,x_k \right) \right] \right\} \hat d  \\
\nonumber
&=  \mp \frac{1}{3\pi ^2\varepsilon _0cd^4}\sum\limits_k R_{0k} \left[
1 - \frac{8d^4\cos \left( \omega _kt \right)}{\left( c^2t^2 - 4d^2
\right)^2} \right. + \frac{2d^2\cos \left( \omega _kt \right) +
d^2\omega _kt\sin \left( \omega _kt \right)}{c^2t^2 - 4d^2} \\
\nonumber
& \qquad - \left[ 2\operatorname{Ci}\left( 2k_kd \right) - \operatorname{Ci}\left( \omega _kt-2k_kd
\right) - \operatorname{Ci}\left( \omega _kt+2k_kd  \right) \right]f\left( k_kd
\right) \\ 
& \qquad \left.  + \left[ 2\operatorname{Si}\left( 2k_kd \right) + \operatorname{Si}\left( \omega
_kt-2k_kd \right) - \operatorname{Si}\left( \omega _kt+2k_kd  \right)-\pi
\right]g\left( k_kd \right) \right]\hat d
\end{align}
\end{widetext}
where the functions $f,g$ are defined by Eq. (\ref{fg}).
It is easy to show that  $\mathbf{F} \to 0$
for $t \to 0$; this is due to the fact that we switch on the interaction at
the initial time $t_0=0$.

\begin{figure}[htbp]
   \centering
   \includegraphics[scale=0.65]{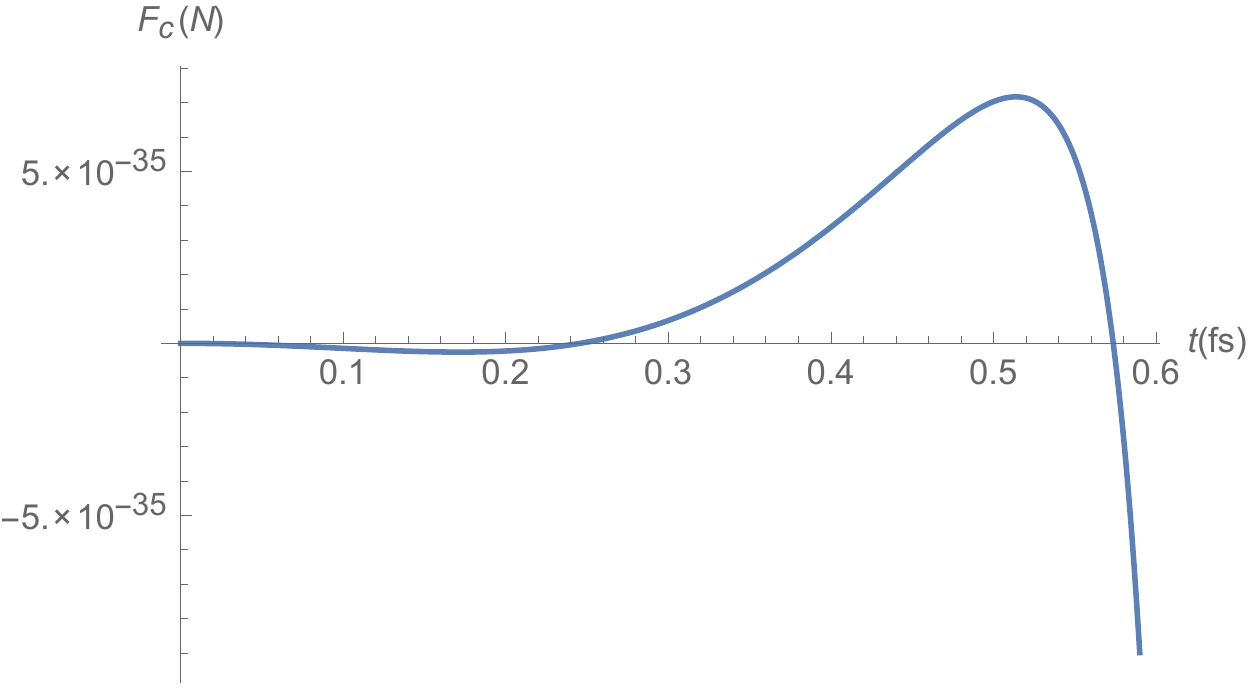}
   \caption{Chiral dynamical Casimir--Polder interaction between an
ground-state dimethyl disulphide and a negative perfect chiral medium
for $d=0.1 \mu\mathrm{m}$ and $t<2d/c=0.67 \mathrm{fs}$.}
    \label{fig2}
   \end{figure}

\begin{figure}[htbp]
   \centering
   \includegraphics[scale=0.65]{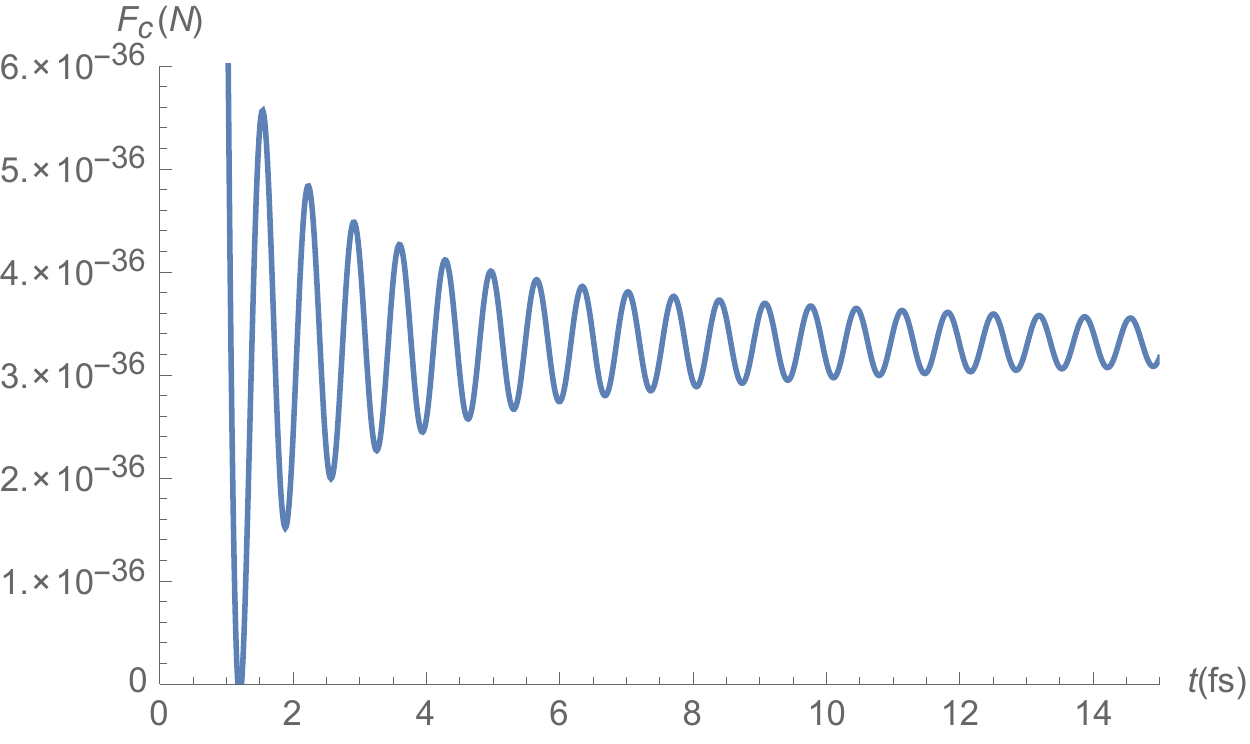}
   \caption{Chiral dynamical Casimir--Polder interaction between an
ground-state dimethyl disulphide and a negative perfect chiral medium
for $d=0.1\,\mu\mathrm{m}$ and $t>2d/c=0.67 \mathrm{fs}$.}
   \label{fig3}
   \end{figure}

For subsequent times,
the force increases exhibiting an oscillatory behaviour in time. Depending on the
time, the force can be attractive and repulsive for a given distance,
contrary to the static case where it has a definite sign.
This is illustrated
in Fig.~\ref{fig2}, where we display
the chiral CP force at fixed distance from the mirror
before the back-reaction time.
   \\
For large times, the force
converges to the static force; this corresponds to a totally dressed
molecule. Fig.~\ref{fig3} shows the dynamics of the chiral force after the
backreaction.
   
To interpret our results, recall that the
CP force is due to the exchange of one virtual photon
between the molecule and the mirror. Different expressions for the
force are needed before and after the backreaction time because the
photon needs a finite time in order to be reflected and absorbed by
the molecule. Note that the force is non-vanishing even before the
backreaction time, because the molecule interacts with the field modes
which incorporate the presence of the conducting wall and hence
instantaneously feels the presence
of the mirror. This is because we are evaluating the force on the atom, which responds to the local field at its position and thus it is immediately influenced by a change of the atom's physical parameters. We expect that, if the force on the conducting wall were evaluated, it would be influenced by a change of the atomic parameters (or by a sudden switching on of the atom-field interaction) only after the causality time $t=d/c$.

The force is divergent on the light cone $t=2d/c$, because in the frequency-integral we include arbitrarily large frequencies: this divergence is
not surprising, being due to the assumption of a point-like molecule (dipole approximation) and to
the idealised nature of the material. A real material is transparent for large frequencies, providing a natural cut-off to regularize the integral; 
moreover, also the inclusion of a finite size of the atom/molecule would provide a natural ultraviolet cutoff given by the appropriate atomic form factor.

\section{\label{Sec5}Conclusions}

Using a dynamical approach, we have obtained the electric, magnetic and
the chiral parts of the time-dependent Casimir--Polder
interaction between an initially bare chiral molecule and a
body at finite temperature, the molecule
initially being prepared in a generic internal state. The force depends
on time because the populations of the excited states of the molecule depend
on time (population-induced dynamics), but also 
because of the initial boundary condition (self-dressing induced
dynamics).

As an example we have considered the particular case of the interaction between an initially
bare ground-state chiral molecule and perfectly reflecting
chiral plate at zero temperature. Here, the force is time-dependent
only because of self-dressing.
The dynamical CP can be attractive or repulsive depending
on time, contrary to the static case where it has a definite sign
for a given distance.

The dynamical interaction is due to the exchange of one virtual photon
between the chiral molecule and the mirror. A characteristic time
scale of the dynamical CP force is the time taken by the
virtual photon to be emitted by the molecule, reflected by the mirror,
and reabsorbed by the molecule (back-reaction time). The dynamical interaction oscillates in time and the scale of these oscillations is related
to the molecular transition frequency. 
The dynamical effect we have considered could in principle be measured by switching on the
interaction between the molecule and the field at the initial time $t_0=0$. Even if this is an idealized situation, it could be approximated by 
the more realistic case of a rapid change of some parameter characterizing the atom--field interaction (strength and/or orientation of the atomic dipole moments, atomic transition frequency by Stark shift, for example) or putting the atom at some distance from the macroscopic body \cite{MPRSV14}.
Another possibility to obtain a dynamical effect could be introducing the mirror at $t_0=0$; however, because in our formalism the presence of the mirror is included in the boundary conditions and not in the system's dynamics, this would require a different approach based on a transformation of field operators and modes relating old and new ones (i.e. before and after switching on the mirror), similarly to the dynamical Casimir effect. 
The use of chiral Rydberg atoms, which have low transition frequencies and large polarizabilities, could be a simpler system to measure the dynamical Casimir force, because in this case the dynamical force evolves on longer timescales \cite{gallagher} ($\tau=10^{-9}s$).

We have shown that the chiral Casimir--Polder interaction shows a discriminatory effect
because it changes sign if the molecule is substituted by its
enantiomer and the dynamical force can hence allow us to
distinguish different enantiomers.

We finally remark that in our approach the medium is considered macroscopically with the
electromagnetic Green tensor. Our model can be generalized to different molecular internal states
and different geometries in which the Green's tensor is known.

\section{Acknowledgement}
RP and LR gratefully acknowledge financial support by the
Julian Schwinger Foundation and by MIUR. SYB is
grateful for support by the DFG (grant BU 1803/3-1) and the Freiburg
Institute for Advanced Studies.

\end{document}